\documentclass[journal]{IEEEtran}

\usepackage{times}
\usepackage{footmisc} 
\usepackage{epsfig}
\usepackage{graphicx}
\usepackage{amsmath}
\usepackage{amssymb}
\usepackage{subfigure}
\usepackage{booktabs}
\usepackage{multirow}
\usepackage[ruled]{algorithm2e}
\usepackage{mathrsfs}
\usepackage{verbatim}
\usepackage{comment}
\usepackage{color}
\usepackage{footnote}
\usepackage[noend]{algpseudocode}
\usepackage{CJKutf8}

\begin{document}

\title{Seeing All From a Few: Nodes Selection Using Graph Pooling for Graph Clustering}

\author{Yiming~Wang,~Dongxia~Chang,~Zhiqiang~Fu
        and~Yao~Zhao,~\IEEEmembership{Senior~Member,~IEEE}
        
\thanks{Y. Wang, D. Chang, Z. Fu and Y. Zhao are with the Institute of Information
Science, Beijing Jiaotong University, Beijing 100044, China, and also
with Beijing Key Laboratory of Advanced Information Science and
Network Technology, Beijing 100044, China (e-mail: wangym@bjtu.edu.cn; dxchang@bjtu.edu.cn zhiqiangfu@bjtu.edu.cn;yzhao@bjtu.edu.cn).}}

\markboth{Journal of \LaTeX\ Class Files,~Vol.~14, No.~8, August~2015}%
{Shell \MakeLowercase{\textit{et al.}}: Bare Demo of IEEEtran.cls for IEEE Journals}

\maketitle

\begin{abstract}
Recently, there has been considerable research interest in graph clustering aimed at data partition using the graph information. However, one limitation of the most of graph-based methods is that they assume the graph structure to operate is fixed and reliable. And there are inevitably some edges in the graph that are not conducive to graph clustering, which we call spurious edges. 
This paper is the first attempt to employ graph pooling technique for node clustering and we propose a novel dual graph embedding network (DGEN), which is designed as a two-step graph encoder connected by a graph pooling layer to learn the graph embedding. In our model, it is assumed that if a node and its nearest neighboring node are close to the same clustering center, this node is an informative node and this edge can be considered as a cluster-friendly edge. Based on this assumption, the neighbor cluster pooling (NCPool) is devised to select the most informative subset of nodes and the corresponding edges based on the distance of nodes and their nearest neighbors to the cluster centers. This can effectively alleviate the impact of the spurious edges on the clustering. Finally, to obtain the clustering assignment of all nodes, a classifier is trained using the clustering results of the selected nodes. Experiments on five benchmark graph datasets demonstrate the superiority of the proposed method over state-of-the-art algorithms.
\end{abstract}


%
\IEEEpeerreviewmaketitle

\section{Introduction}

\begin{figure*}[t]
	\centering
	\includegraphics[width=16cm]{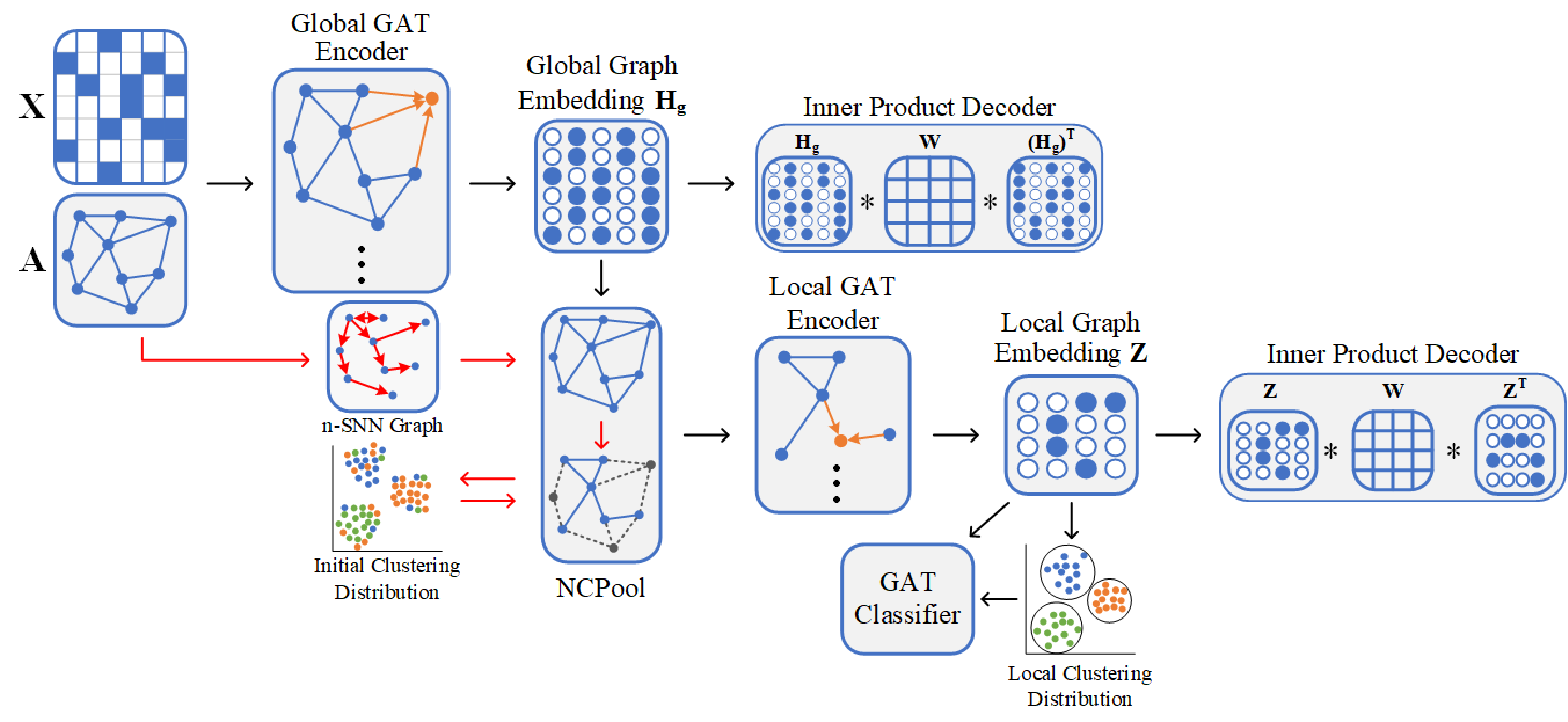}
	\caption{The framework of the proposed DGEN. Our network consists of a two-step graph embedding, namely learning the representations of all nodes based on the original graph and learning the representations of selected nodes based on the pooled local graph.}
\end{figure*}

\IEEEPARstart{W}{ith} the advance of information technology, non-Euclidean domain data can be readily acquired in many domains, such as human pose\cite{journals/tnn/ZhangXTT20}, protein structure\cite{conf/nips/YingY0RHL18} and social networks\cite{journals/kbs/DhelimAN20}. With the emergence of non-Euclidean data, graph clustering methods\cite{journals/pami/ShiM00,journals/tkde/WangYL20} arise at the historic moment, and they are able to learn the adjacency between nodes to boost clustering performance. 
These conventional methods, however, proved to be incomplete in mining nodes relationship for their over-reliance on prior knowledge.

To mine non-Euclidean graph information, researchers have devised graph neural networks (GNNs) \cite{conf/iclr/KipfW17,conf/iclr/VelickovicCCRLB18} capable of encoding both graph structure and node characteristics for node latent representation. GNNs have successfully expanded deep learning techniques to non-Euclidean graph data with remarkable achievement made in multiple graph tasks, such as graph classification\cite{conf/nips/LiC0T20} and visual question answering\cite{conf/iccv/LiGCL19}. Thanks to the properties of graph convolution, many GNN-based graph clustering methods\cite{conf/ijcai/WangPHLJZ19,conf/www/FanWSLLW20} have been proposed. These methods typically construct adjacency matrices based on the k-NN graph\cite{conf/www/Bo0SZL020} or attributed graph\cite{conf/ijcai/ChengWTXG20}, and learn graph embedding using graph autoencoder (GAE)\cite{journals/corr/KipfW16a}. Finally, conventional clustering algorithms such as K-means\cite{conf/soda/ArthurV07} are applied based on the learned graph embedding. For all the promising results these methods may generate, the spurious connections in the initial graph may deteriorate the clustering performance. 

In recent years, graph pooling\cite{conf/nips/DefferrardBV16} emerged to learn abstract representation of the input graph by summarizing local components and discarding redundant information. With growing interest in graph pooling, some improved methods have been proposed. Current graph pooling methods can be divided into two classes: differentiable graph pooling methods\cite{conf/nips/YingY0RHL18, kdd/0001WAT19, conf/icml/BianchiGA20, 9211767} and sorting-based graph pooling methods\cite{conf/icml/GaoJ19, nips/LiC0T20, 9366345, 9266105}. 

Differentiable pooling was first proposed by DiffPool\cite{conf/nips/YingY0RHL18}. It can learn a assignment matrix at each GNN layer and generate hierarchical representation of graphs. To effectively capture the graph substructure, ASAP\cite{conf/aaai/RanjanST20} propose a sparse pooling operator capable of capturing local subgraph information hierarchically. HiGPool\cite{9141427} designs a pooling module that can capture the spatial layouts of points to learn the hierarchical features adequately and applies it to point cloud segmentation. Differentiable pooling works well in several tasks, but suffers from two main disadvantages in node clustering: (a) there is no clear mapping between the nodes of the previous and subsequent layers, which is crucial for clustering tasks; (b) these methods have a quadratic storage complexity and the number of its parameters is dependent on the number of nodes.

Sorting-based graph pooling learns a projection vector that is applied to each node feature to obtain a score. Graph U-Net\cite{conf/icml/GaoJ19} uses a top-$k$ choice of nodes for their gPool layer capable of learning a node score and dropping low score nodes. \cite{journals/corr/abs-1811-01287} applied this to graph classification and proposed TopKPool which achieved comparable performance with DiffPool. SAGPool\cite{conf/icml/LeeLK19} takes graph topology into account and devises a self-attention graph pooling method for GNNs in the context of hierarchical graph pooling. To preserve diverse representative nodes in different neighborhoods, iPool\cite{9392315} leverages the proposed neighborhood information gain criterion to select informative nodes in each neighborhood. Sorting-based methods do not break the mapping from the nodes of the previous layer to the next layer. But the selected nodes may not be distributed across the essential areas in the graph, which may cause the absence clusters in clustering tasks.

The methods mentioned above generally utilize localized node features and connecting relationship to obtain a smaller graph representation. However, all these graph pooling methods focus on graph representation rather than node representation. Therefore, they can be used to select the nodes and edges that best characterize the graph. But it is difficult to apply them directly to remove spurious edges in graph clustering.

In this work, to achieve more efficient and accurate nodes clustering on graph, we design a new dual graph embedding network (DGEN), which learns global graph embedding and local graph embedding step by step, with the two processes connected by a graph pooling layer. To reduce the impact of spurious edges on graph clustering, we design a novel neighbor cluster pooling(NCPool), which selects the most informative subset of nodes and the corresponding edges based on the distance of nodes and their nearest neighbors to the cluster centers. And the fundamental idea of the pooling is that nodes and their nearest neighbors should be close to the same clustering center. To obtain the nearest neighbor of each node in the attribute graph, we compute the shared nearest neighbor(SNN)\cite{jarvis1973clustering} similarity on the graph. Then conventional clustering algorithm K-means is performed on the selected node representations to obtain the cluster assignment of the selected nodes. Finally, the selected nodes and their labels are used to train a classifier so as to obtain the final clustering assignments.

Our major contributions can be summarized as follows:
\begin{itemize}
    \setlength{\itemsep}{0pt}
    \setlength{\parsep}{0pt}
    \item We propose a dual graph embedding network for clustering on the graph-structured data. To the best of our knowledge, it is the first time to employ graph pooling to select informative nodes for graph clustering.
    \item A novel neighbor clustering pooling (NCPool) is devised to capture local graph information, which can effectively enhance the robustness of graph clustering for the spurious edges.  
    
    \item Extensive experiments on five benchmark graph datasets show that our DGEN outperforms state-of-the-art graph clustering methods.
\end{itemize}

\section{Proposed Methodology}

In order to improve the robustness of graph clustering for the spurious edges, we propose NCPool which uses node features and graph topology to choose informative nodes and corresponding edges that are more suitable for clustering. And a novel dual graph embedding network(DGEN) based on this layer is proposed, which is shown in Fig.~1. Firstly, a graph autoencoder is applied to learn the global graph embedding on which NCPool is performed. Local graph embedding is then learned based on the selected nodes and edges. Finally, K-means is employed for the selected node representations to obtain the local clustering assignments and the local clustering assignments are used as labels to train a classifier that can obtain the final clustering assignments. In the following, we will describe our proposed model in detail.

Let a graph be represented by a triple $G = \{\mathcal{V}, \mathcal{E}, X\}$  with $N = |\mathcal{V}|$ nodes and $|\mathcal{E}|$ edges. The $d$-dimensional feature of each node $v_i \in \mathcal{V}$ is denoted by $x_i$, and $X \in \mathbb{R}^{N \times d}$ denotes the node feature matrix. The topological structure of $G$ can be represented by an adjacency matrix $A$, where $A_{i,j} = 1$ if $(v_i, v_j) \in \mathcal{E}$; otherwise $A_{i,j} = 0$.

\subsection{NCPool}

In graph embedding, spurious edges can reduce the discriminability of the learned node representation, which in turn affects the performance of graph clustering. To reduce the impact of spurious edges on graph clustering, we propose a novel neighbor cluster pooling(NCPool) to select the nodes that are close to the same clustering centre as their neighbours.

Here, a pooling function $S(\cdot)$ is introduced to select the subset $\Phi \subset V$ containing $pN$ nodes. And the pooling problem can be formulated as
\begin{equation}
    \min_{\Phi \subset V}\ S(\Phi),\ {\rm subject\ to}\ |\Phi| = pN,
\end{equation}
where $p \in (0,1)$ denotes the ratio of selected nodes, and $S(\Phi)$ is designed based on the distance of the nodes and their nearest neighbors to cluster centers. In order to select nodes that fit the clustering distribution better, we calculate the cluster centers $c$ by K-means\cite{conf/soda/ArthurV07}. There are two main advantages of using K-means: (1) it is fast and efficient, with little impact on the speed of model training; (2) in general, graph embedding obeys the Gaussian distribution, which satisfies the assumption of K-means.

The distance between the node and each cluster center is first calculated and we can obtain the distance between the node and the closest cluster center
\begin{equation}
    \sum_{v_i \in \Phi}||H_{v_i} - H_{c_{i,j}}||_2^2 ,
\end{equation}
where $H_{v_i}$ is the node representation of $v_i$ fed into the pooling layer. And $c_{i,j}$ is the closest cluster center to $v_i$.

To find the nearest neighbor of each node on the attributed graph, we use a modified Shared Nearest Neighbor(SNN) \cite{jarvis1973clustering} to define the similarity of the nodes on the graph. And the modified SNN similarity is written as
\begin{equation}
sim(i,j)=\left\{\begin{matrix}
0 ,& (v_i, v_j) \notin \mathcal{E},\\ 
|\mathcal{N}(i)\cap \mathcal{N}(j)|, & (v_i, v_j) \in \mathcal{E}.
\end{matrix}\right.
\end{equation}
where $\mathcal{N}(*)$ denotes the neighboring nodes of $v_*$. After getting the similarity of all other points to $v_*$, the most similar node is selected as the nearest neighbor of $v_*$.

Then the final $S(\Phi)$ can be calculated by

\begin{equation}
    S(\Phi) =  \sum_{v_i \in \Phi}(||H_{v_i} - H_{c_{i,j}}||_2^2 + ||H_{v_{in}} - H_{c_{i,j}}||_2^2),
\end{equation}
where $v_{in}$ denotes the nearest neighbor of $v_i$.

After obtaining the score of all nodes, we apply the node selection method proposed by \cite{journals/corr/abs-1811-01287,conf/icml/GaoJ19}, which retains a portion of nodes of the input graph even when graphs of varying sizes and structures are inputted. The top $\lceil kN \rceil$ nodes are selected based on the value of $S$.
\begin{equation}
    idx = {\rm top}_k(S, \lceil kN \rceil),
\end{equation}
And an input graph is processed by the following operation.
\begin{equation}
    H' = H_{idx,:}\odot (\mathbf{a}\mathbf{1}^T),\ \  A' = A_{idx,idx}
\end{equation}
where $\mathbf{a}$ denotes an affinity vector consisting of the score for the selected nodes, and $\mathbf{1}$ is an all-one vector. $H_{idx,:}$ is the row-wise indexed feature matrix, and $A_{idx,idx}$ is the row-wise and col-wise indexed adjacency matrix. 

\subsection{Dual graph embedding network}

In this section, the proposed dual graph embedding network(DGEN) which learns the graph embedding of the subset of nodes and edges for graph clustering is described in detail. In our model, the global graph embedding $H_g$ is first learned via the Global GAT Encoder. Then, $H_g$ is fed into our NCPool to obtain informative nodes and edges, and the final embedding $Z$ is obtained by the Local GAT Encoder. Finally, the local clustering process is conducted on the informative embedding Z, and the labels obtained is used to train a GAT classifier which is used to obtain the final clustering assignments for all the nodes.  

In our DGEN, graph attention layer(GAT)\cite{conf/ictai/SalehiD20} with multi-head is used to learn the representation of neighboring nodes adaptively. Let $h^{(l)}_i$ be the latent representation of the node $v_i$ learned by the $l$-th layer, and the GAT with $m$-head can be expressed as
\begin{equation}
    h^{(l+1)}_i = {\rm concat}_1^m(\phi(\sum_{j \in \mathcal{N}(i)} \alpha^m_{i,j}W^{(l)m}h^{(l)}_{j})),
\end{equation}
where $\mathcal{N}(i)$ denotes the neighboring nodes of $v_i$, ${\rm concat}_1^m(\cdot)$ means a vector stitching from the first one to $m$-th. And the attention coefficients $\alpha_{i,j}$ can be computed as
\begin{equation}
    \alpha_{i,j} =
\frac{
\exp\left(\mathrm{LeakyReLU}\left(\mathbf{a}^{\top}
[Wh_i \, \Vert \, Wh_j]
\right)\right)}
{\sum_{k \in \mathcal{N}(i) \cup \{ i \}}
\exp\left(\mathrm{LeakyReLU}\left(\mathbf{a}^{\top}
[Wh_i \, \Vert \, Wh_k]
\right)\right)}.
\end{equation}

Each GAT encoder includes two GAT layers. And in order to guide the GAT encoders to learn a comprehensive node representation, an inner product decoder is applied to reconstruct the graph data $\hat{A}$ from the global embedding $Z$.
\begin{equation}
    \hat{A}_{i,j} = {\rm sigmoid} (z_i\cdot z_j^T) 
\end{equation}
where $\hat{A}_{i,j}$ denotes the adjacency of node $v_i$ and node $v_j$ in the reconstructed graph $\hat{A}$.
Then, we minimize the reconstruction error by measuring the difference between $A$ and $\hat{A}$.
\begin{equation}
    L_r = \sum_i^N loss(A_{i,j}, \hat{A}_{i,j} )
\end{equation}

The above part merely aims to obtain the representation of selected nodes, but cannot guarantee that the obtained representation is cluster-friendly. Inspired by the good properties of ‘KL divergence’ based clustering\cite{icml/XieGF16, conf/ijcai/WangPHLJZ19}, we develop a self-optimizing training module as a solution to overcome this difficulty. Specifically, for the local graph embedding $Z$, if its cluster center is denoted by $\mu_j$, the loss function can be written as 
\begin{equation}
L_c = KL(P||Q) = \sum_i\sum_jp_{ij}\rm{log}\frac{p_{ij}}{q_{ij}}
\end{equation}
In Eq.(11), $q_{ij}$ can be calculated using the Student’s t-distribution\cite{jmlr/laurens08} as follows
\begin{equation}
q_{ij} = \frac{(1+||z_i - \mu_j||^2)^{-1}}{\sum_k(1+||z_i - \mu_k||^2)^{-1}}
\end{equation}
where $z_i$ is the $i$-th row of $Z$, $\mu_j$ is initialized by K-means on the pre-trained representation. $q_{ij}$ can be seen as the probability of assigning node $z_i$ to cluster $j$. $Q = [q_{ij}]$ is the distribution of the assignments of all nodes. $P$ is the target distribution of $Q$ and can be calculated as 
\begin{equation}
p_{ij} = \frac{q_{ij}^2/\sum_i q_{ij}}{\sum_k(q_{ik}^2/\sum_i q_{ik})}
\end{equation}
By minimizing the KL divergence loss between distributions $P$ and $Q$, a more cluster-friendly representation can be produced according to the high confidence predictions. Thus, the total objective function of DGEN is defined as:
\begin{equation}
L = L_r + \lambda L_c
\end{equation}
where $\lambda$ is the hyperparameter balancing these two losses.

Now, we can obtain the clustering assignments of the selected nodes. However, our goal is to obtain the cluster assignment of all nodes. 
To obtain the global clustering assignments, a GAT classifier is trained using the selected nodes, corresponding edges and local clustering assignments. And the final clustering assignments can be obtained by inputting all nodes and selected edges to the classifier. 
There is a notable advantage for our model to use the classifier to obtain the 
final clustering assignments instead of obtaining them directly based on the 
adjacent relations. Spurious edges between discard nodes and selected nodes can 
lead to inaccurate classification.

\section{Experiments}

\subsection{Experimental Settings}

\begin{table}[t]
\footnotesize
\centering
\caption{Benchmark Graph Datasets}
\vspace{-0.5em}
\begin{tabular}{lcccc}
\toprule
Dataset  & Clusters & Nodes & Features & Links \\
\midrule
Cora\footnotemark[1]     & 7        & 2708  & 1433     & 5429  \\
CiteSeer\footnotemark[1]  & 6        & 3327  & 3703     & 4732  \\
PubMed\footnotemark[1]    & 3        & 19717 & 500      & 44338 \\
DBLP\footnotemark[2]      & 4        & 4058  & 334      & 7056  \\
ACM\footnotemark[3]       & 3        & 3025  & 3703     & 26256 \\
\bottomrule
\end{tabular}
\end{table}
\footnotetext[1]{https://linqs.soe.ucsc.edu/data}
\footnotetext[2]{https://dblp.uni-trier.de}
\footnotetext[3]{http://dl.acm.org/}

\paragraph{Datasets}
In order to evaluate the effectiveness of our proposed method, we conduct experiments on five citation networks widely-used to assess the attributed graph analysis. These datasets are summarized in Table~\uppercase\expandafter{\romannumeral1}. 

\paragraph{Evaluation Metrics}
Three standard evaluation metrics: Accuracy (ACC), Normalized Mutual Information (NMI) and Adjusted Rand Index (ARI) are used to evaluate the performance. In all cases, the values of these metrics range between 0 and 1, where higher values correspond to better clustering performance.

\paragraph{Baseline Methods}
In order to verify the effectiveness of the proposed DGEN, we compare it with a total of 12 clustering algorithms, including methods that use only node attribute or graph structure, and both.
\begin{itemize}
    \setlength{\itemsep}{0pt}
    \setlength{\parsep}{0pt}
    \setlength{\parskip}{0pt}
    \item \textbf{Using node attribute:} Kmeans\cite{conf/soda/ArthurV07};
    \item \textbf{Using graph structure:} Spectral clustering\cite{conf/nips/NgJW01}, denoising autoencoder for graph embedding (DNGR)\cite{conf/aaai/CaoLX16}, and modularized nonnegative matrix factorization (M-NMF)\cite{wang2017community};
    \item \textbf{Using both node attribute and graph structure:} Graph auto-encoders (GAE) and variational graph auto-encoders (VGAE)\cite{journals/corr/KipfW16a}, marginalized graph autoencoder (MGAE)\cite{conf/cikm/WangPLZJ17}, adversarial regularized graph autoencoder (ARGAE) and adversarial variational regularized graph autoencoder (ARVGAE)\cite{conf/ijcai/PanHLJYZ18}, deep attentional embedding graph clustering (DAEGC)\cite{conf/ijcai/WangPHLJZ19}, graph attention auto-encoders (GATE)\cite{conf/ictai/SalehiD20}, and distribution-induced bidirectional generative adversarial network (DBGAN)\cite{conf/cvpr/ZhengZZL0020}.
\end{itemize}

\begin{table}[t]
\centering
\caption{The detailed configuration of the proposed DGEN}
\vspace{-0.5em}
\resizebox{8cm}{!}{
\begin{tabular}{lccc}
\toprule
Network module                      & Layer & Output dimension   & Heads             \\
\midrule
\multirow{2}{*}{Global GAT Encoder} & GAT-1 & 256                & 8(4 for CiteSeer) \\
                                    & GAT-2 & 16                 & 1                 \\
\midrule
\multirow{2}{*}{Local GAT Encoder}  & GAT-1 & 16                 & 1                 \\
                                    & GAT-2 & 10                 & 1                 \\
\midrule
\multirow{2}{*}{GAT classifier}     & GAT-1 & 8                  & 8                 \\
                                    & GAT-2 & Number of clusters & 1(8 for PubMed)   \\ 
\bottomrule
\end{tabular}}
\end{table}

\paragraph{Implementation Details}
In our experiments, the ratio of selected nodes of NCPool is set to 0.6, and the hyperparameter $\lambda$ is set to 10. Moreover, there are three main network modules in our DGEN. A detailed configurations of these network modules are given in Table~\uppercase\expandafter{\romannumeral2}. Note that, the Local GAT Encoder is constructed with a 10-neuron embedding layer for all datasets, which is much smaller than the commonly used 64-neuron. Adam optimizer is used with learning rate  $\alpha = 0.001$ for GAT encoders, and $\alpha = 0.0005$ for GAT classifier.

\subsection{Comparison of Clustering Performance}

\begin{table*}[!t]
\centering
\caption{Performance comparisons of various methods on five benchmark datasets}
\resizebox{\textwidth}{!}{
\begin{tabular}{c|c|ccc|ccc|ccc|ccc|ccc}
\toprule
\multirow{2}{*}{Method} & \multirow{2}{*}{Info.} & \multicolumn{3}{c|}{Cora} & \multicolumn{3}{c|}{CiteSeer} & \multicolumn{3}{c|}{PubMed}& \multicolumn{3}{c|}{DBLP}& \multicolumn{3}{c}{ACM} \\ 
                        &        & ACC    & NMI    & ARI    & ACC      & NMI     & ARI     & ACC     & NMI     & ARI  & ACC    & NMI    & ARI    & ACC      & NMI     & ARI  \\ 
\midrule
K-means                 & X      & 0.500& 0.317& 0.239& 0.544& 0.312& 0.285& 0.580& 0.278& 0.246& 0.387& 0.115& 0.070 &0.673& 0.324& 0.306\\
\midrule
Spectral                & G      & 0.398& 0.297& 0.174& 0.308& 0.090& 0.082& 0.496& 0.147& 0.098& 0.295& 0.017& 0.014& 0.351& 0.133& 0.174\\
DNGR                    & G      & 0.419& 0.318& 0.142& 0.326& 0.180& 0.043& 0.468& 0.153& 0.059& 0.496& 0.187& 0.154& 0.551& 0.213& 0.266\\
M-NMF                   & G      & 0.423& 0.256& 0.161& 0.336& 0.099& 0.070& 0.470& 0.084& 0.058& 0.503& 0.110& 0.187& 0.604& 0.189& 0.291\\
\midrule
GAE                     & G\&X   & 0.530& 0.397& 0.293& 0.380& 0.174& 0.141& 0.632& 0.249& 0.246& 0.612& 0.308& 0.220& 0.845& 0.554& 0.595\\
VGAE                    & G\&X   & 0.592& 0.408& 0.347& 0.392& 0.163& 0.101& 0.619& 0.216& 0.201& 0.586& 0.269& 0.179& 0.841& 0.530& 0.577\\
MGAE                    & G\&X   & 0.684& 0.511& 0.448& 0.661& \textcolor{blue}{0.412}& \textcolor{blue}{0.414}& 0.593& 0.282& 0.248& 0.627 & 0.313& 0.237 & 0.872 & 0.617 & 0.608       \\
ARGAE                   & G\&X   & 0.640& 0.449& 0.352& 0.573& 0.350& 0.341& 0.681& 0.276& 0.291 & 0.379& 0.107& 0.105& 0.825& 0.479& 0.544\\
ARVGAE                  & G\&X   & 0.638& 0.450& 0.374& 0.544& 0.261& 0.245& 0.513& 0.117& 0.078 & 0.402& 0.114& 0.106& 0.867& 0.563& 0.640\\
DAEGC                   & G\&X   & 0.704& 0.528& 0.496& \textcolor{blue}{0.672}& 0.397& 0.410& 0.671& 0.266& 0.278 & 0.621& 0.325& 0.210& 0.869& 0.562& 0.594 \\
GATE                    & G\&X   & 0.658& 0.527& 0.451& 0.616& 0.401& 0.381& 0.673& \textcolor{blue}{0.322}& 0.299 & 0.599& 0.317& 0.200& 0.841& 0.503& 0.549     \\
DBGAN                   & G\&X   & \textcolor{blue}{0.748}& \textcolor{blue}{0.560}& \textcolor{blue}{0.540}& 0.670& 0.407& \textcolor{blue}{0.414}& \textcolor{blue}{0.694}& \textcolor{red}{0.324}& \textcolor{red}{0.327}& \textcolor{blue}{0.657}&\textcolor{blue}{0.337}&\textcolor{blue}{0.237}&\textcolor{blue}{0.889}&\textcolor{red}{0.635}&\textcolor{blue}{0.699} \\
\midrule
DGEN                    & G\&X   & \textcolor{red}{0.771}& \textcolor{red}{0.576}& \textcolor{red}{0.566}& \textcolor{red}{0.686}& \textcolor{red}{0.430}& \textcolor{red}{0.445}& \textcolor{red}{0.695}& 0.297& \textcolor{blue}{0.320} & \textcolor{red}{0.667}&\textcolor{red}{0.340}&\textcolor{red}{0.306}&\textcolor{red}{0.890}&\textcolor{blue}{0.629}&\textcolor{red}{0.708}     \\
\bottomrule
\end{tabular}}

\end{table*}

\paragraph{Evaluation Metrics} The results on the five benchmark datasets are reported in Table~\uppercase\expandafter{\romannumeral3}, where the top value is highlighted in red font and the second best in blue. $X$, $G$, and $G$\&$X$ indicate that the method uses only node attribute, graph structure, or both attribute and structure information, respectively. According to these results, we have the following observations: (1) our proposed method DGEN outperforms the competing methods on these benchmark datasets in most cases. For example, DGEN shows significant improvement over the base method GATE in ACC and ARI on Cora by a margin of around 11.3\% and 11.5\%, respectively; (2) methods using both node attribute and graph structure consistently obtain better performances than methods using only one type of information, which demonstrates that both the graph structure and node attribute contain useful information for graph clustering; (3) DGEN achieves competitive results than other GNN-based method, which indicates the feasibility of applying graph pooling to graph clustering and the effectiveness of our proposed NCPool. The second best baseline (DBGAN) indeed outperforms our DGEN in several cases. This, however, is achieved at the cost of a higher representation dimension.  
\begin{figure*}[!t]
	\centering
	\subfigure[Cora(raw)]{	
		\label{fig2:a} 
		\includegraphics[width=3.3cm]{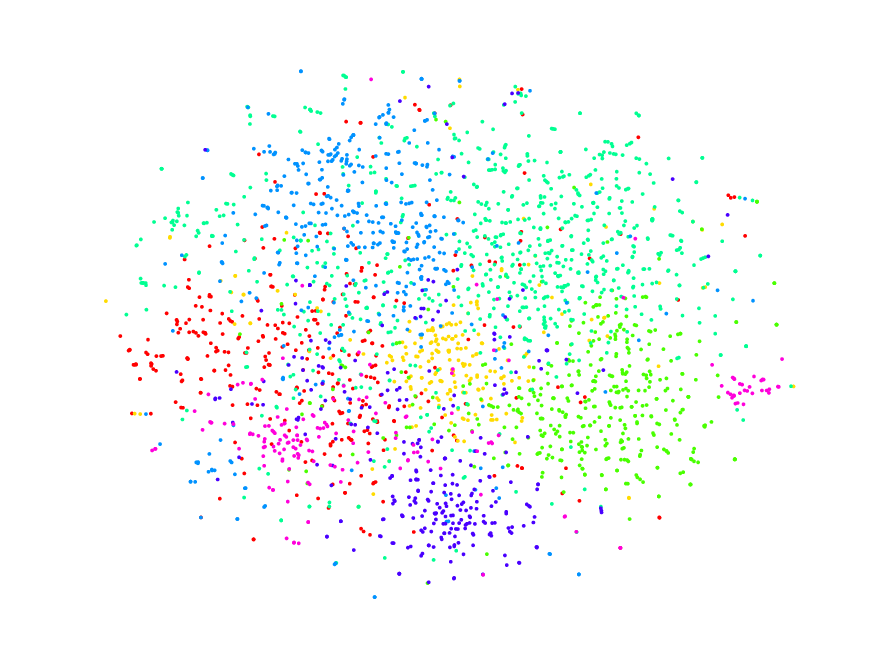}}
	\subfigure[Cora(GAE)]{	
		\label{fig2:b}
		\includegraphics[width=3.3cm]{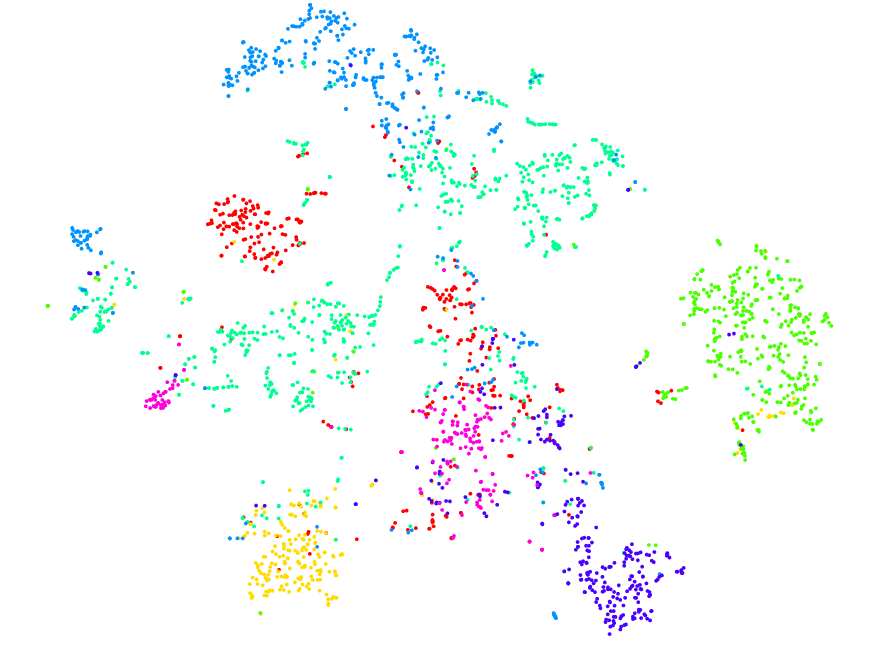}}
	\subfigure[Cora(GATE)]{	
		\label{fig2:c} 
		\includegraphics[width=3.3cm]{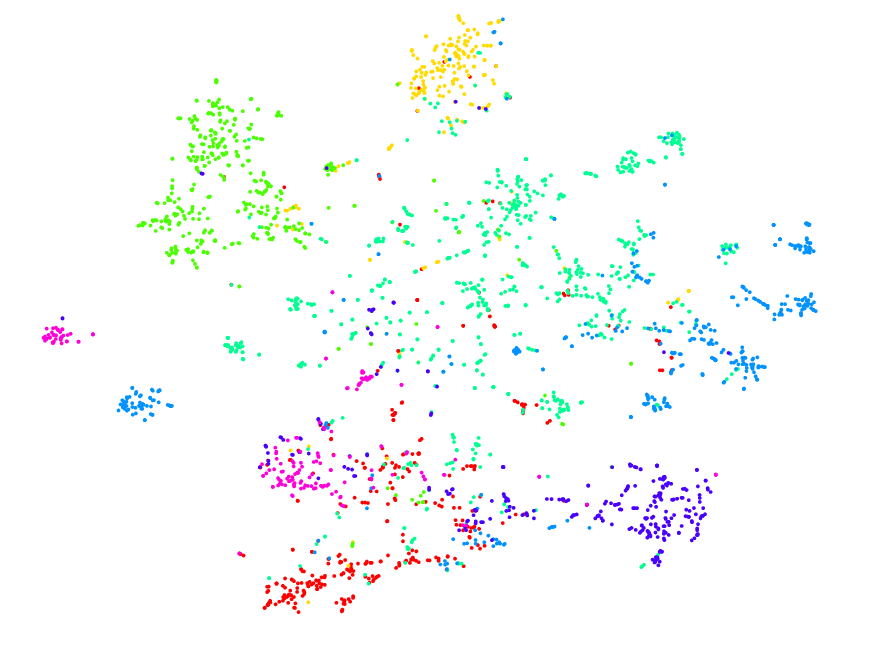}}
	\subfigure[Cora(DGEN-G)]{	
		\label{fig2:d}
		\includegraphics[width=3.3cm]{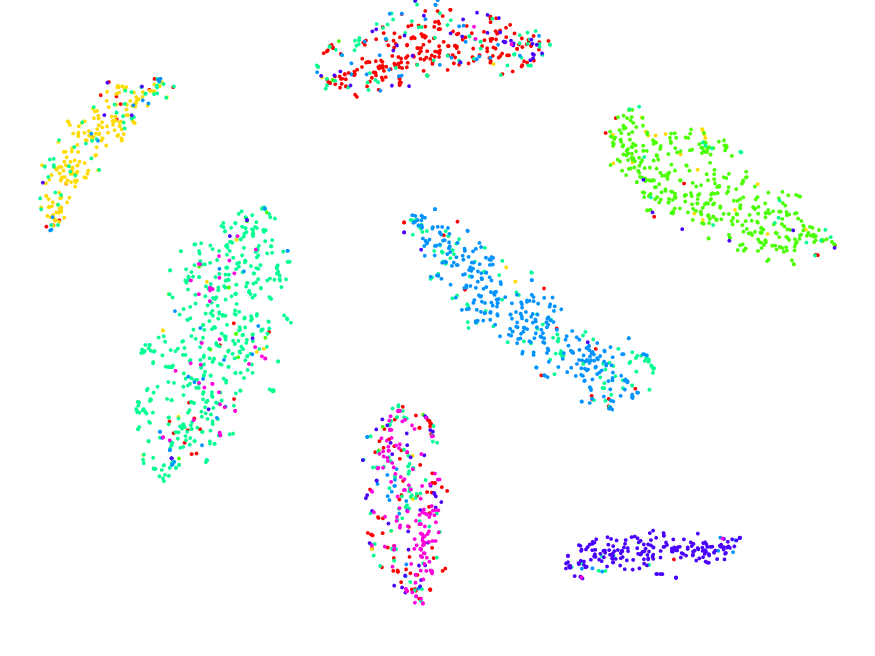}}
	\subfigure[Cora(DGEN-L)]{
		\label{fig2:e} 
		\includegraphics[width=3.3cm]{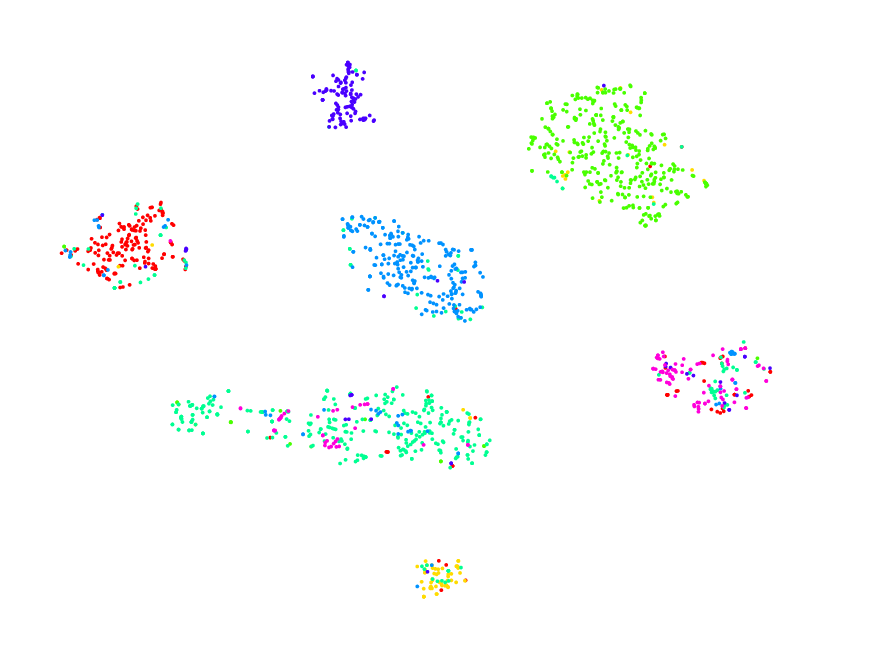}}
	\vfill
	\subfigure[PubMed(raw)]{
		\label{fig2:f} 
		\includegraphics[width=3.3cm]{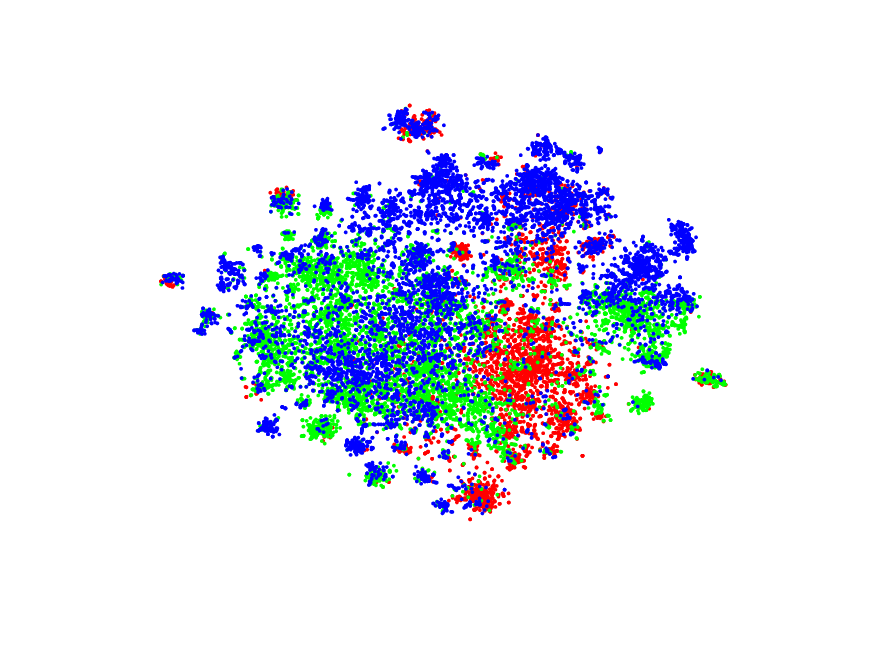}}
	\subfigure[PubMed(GAE)]{	
		\label{fig2:G} 
		\includegraphics[width=3.3cm]{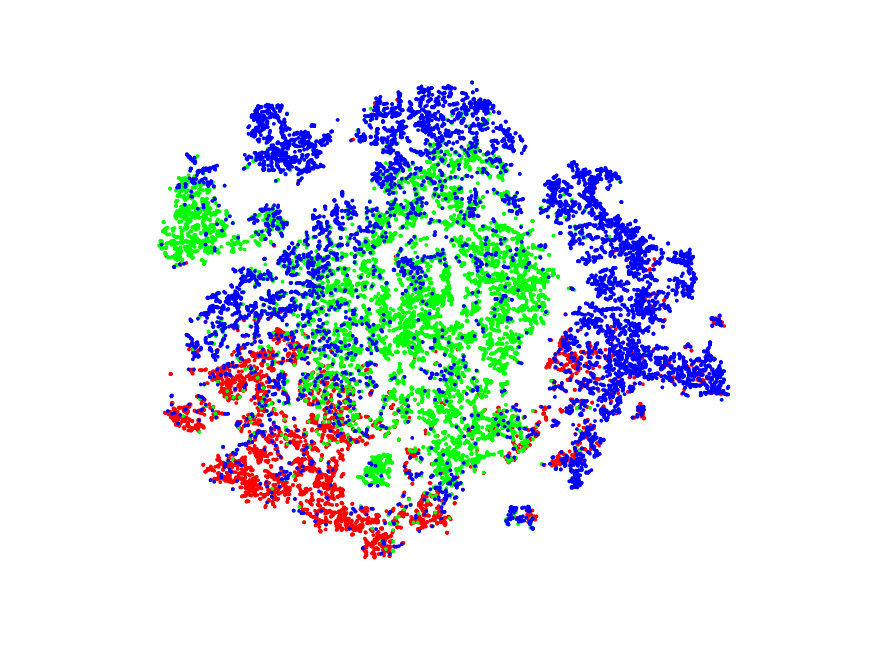}}
	\subfigure[PubMed(GATE)]{	
		\label{fig2:H}
		\includegraphics[width=3.3cm]{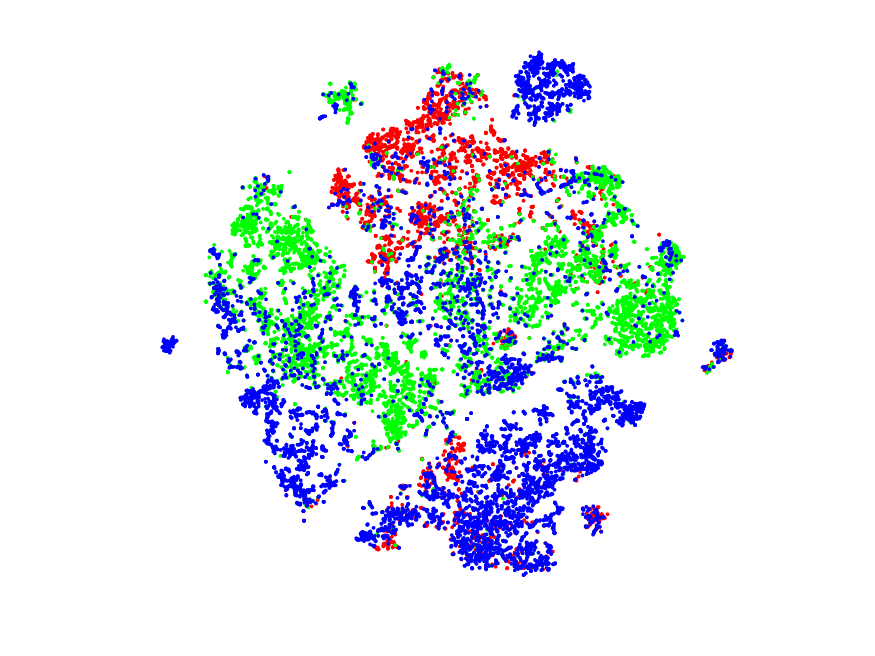}}
	\subfigure[PubMed(DGEN-G)]{	
		\label{fig2:i}
		\includegraphics[width=3.3cm]{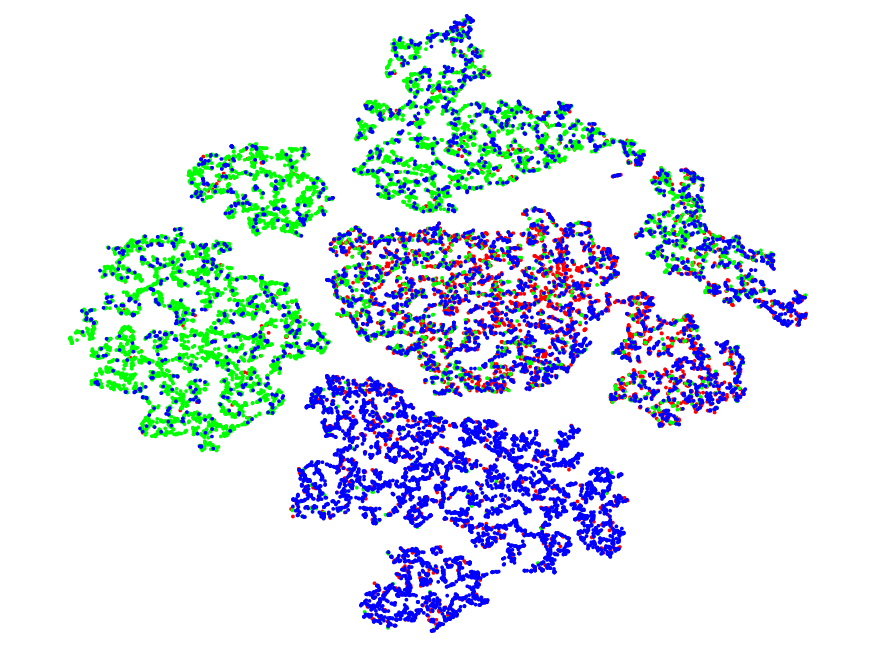}}
	\subfigure[PubMed(DGEN-L)]{	
		\label{fig2:j}
		\includegraphics[width=3.3cm]{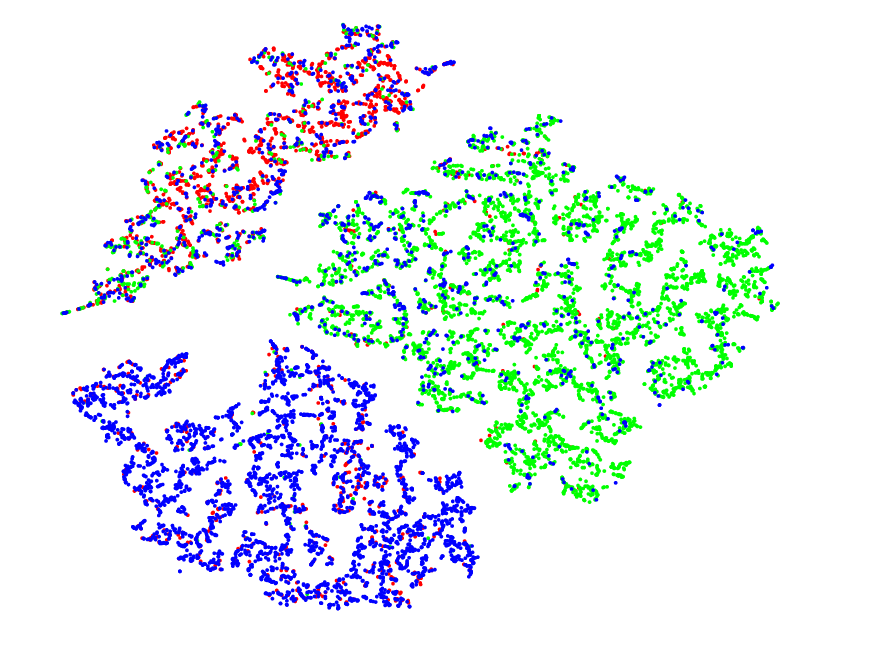}}
	\caption{t-SNE visualizations of representations learned by various methods on Cora (top row) and PubMed (bottom row).}
	\label{fig2} 
\end{figure*}

\paragraph{Visualization} In order to show the superiority of the representation obtained by our method, we visualize the results of different methods on Cora and PubMed with t-SNE\cite{jmlr/laurens08}. The results are shown in Fig.~2, where DGEN-G and DGEN-L denote the representation of global graph embedding $H_g$ and local graph embedding $Z$, respectively. As can be seen from Fig.~2, our method can recover better cluster structure of data since it has smaller intra-cluster scatter and larger inter-cluster scatter. Our proposed NCPool method also proves effective since representation from DGEN-L has fewer intra-cluster error points than DGEN-G.

\begin{table}[t]\small
\caption{Clustering results of different graph pooling methods}
\vspace{-1em}
\begin{center}
\resizebox{8.5cm}{!}{
\begin{tabular}{c|ccc|ccc|ccc}
\toprule
\multirow{2}{*}{Method} & \multicolumn{3}{c|}{Cora} & \multicolumn{3}{c|}{CiteSeer} & \multicolumn{3}{c}{PubMed} \\ 
                        & ACC & NMI & ARI & ACC & NMI & ARI & ACC & NMI & ARI  \\ 
\midrule
TopKPool             &0.482&0.296&0.178&0.571&0.354&0.310&0.544&0.165&0.166 \\
SAGPool              &0.654&0.525&0.478&0.602&0.355&0.353&0.593&0.196&0.136 \\
NCPool               &\textbf{0.771}&\textbf{0.576}&\textbf{0.566}&\textbf{0.686}&\textbf{0.430}&\textbf{0.445}&\textbf{0.695}&\textbf{0.297}&\textbf{0.320} \\
\bottomrule
\end{tabular}}
\end{center}  
\end{table}

\paragraph{Clustering capability of different graph pooling methods} We have mentioned why current graph pooling methods are difficult to be applied to select nodes for clustering. To verify it, the NCPool is replaced with other two graph pooling methods and the results are given in Table~\uppercase\expandafter{\romannumeral4}. The experimental results demonstrate that NCPool can select nodes that are more cluster-friendly compared with the other two methods.

\begin{figure}[t]
	\centering
	\subfigure[ACC]{	
		\label{fig3:a} 
		\includegraphics[width=4.15cm]{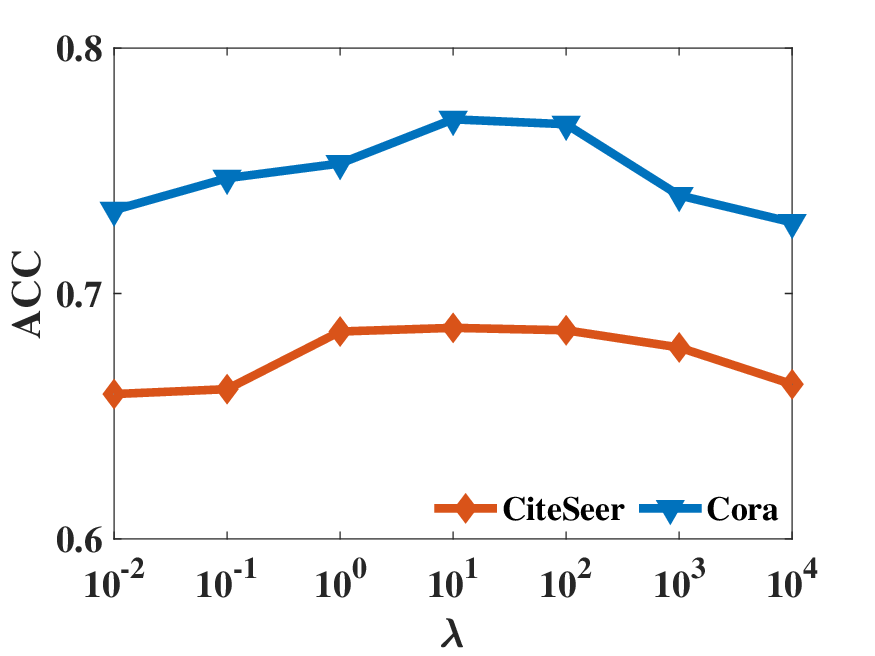}}
	\subfigure[NMI]{	
		\label{fig3:b}
		\includegraphics[width=4.15cm]{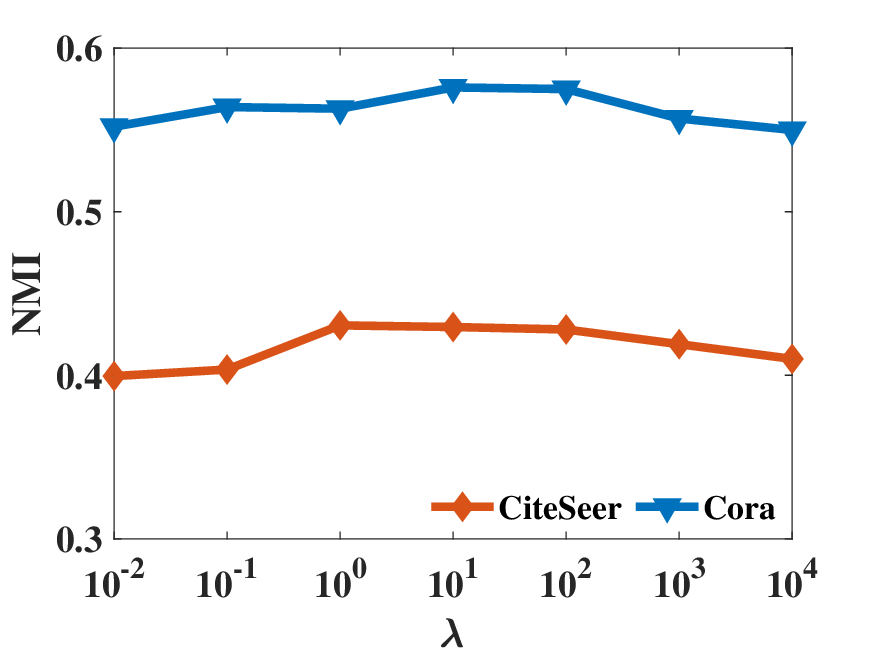}}
	\caption{Clustering results with different $\lambda$.}
		\label{fig3} 
\end{figure}

\begin{figure}[t]
	\centering
	\includegraphics[width=9cm]{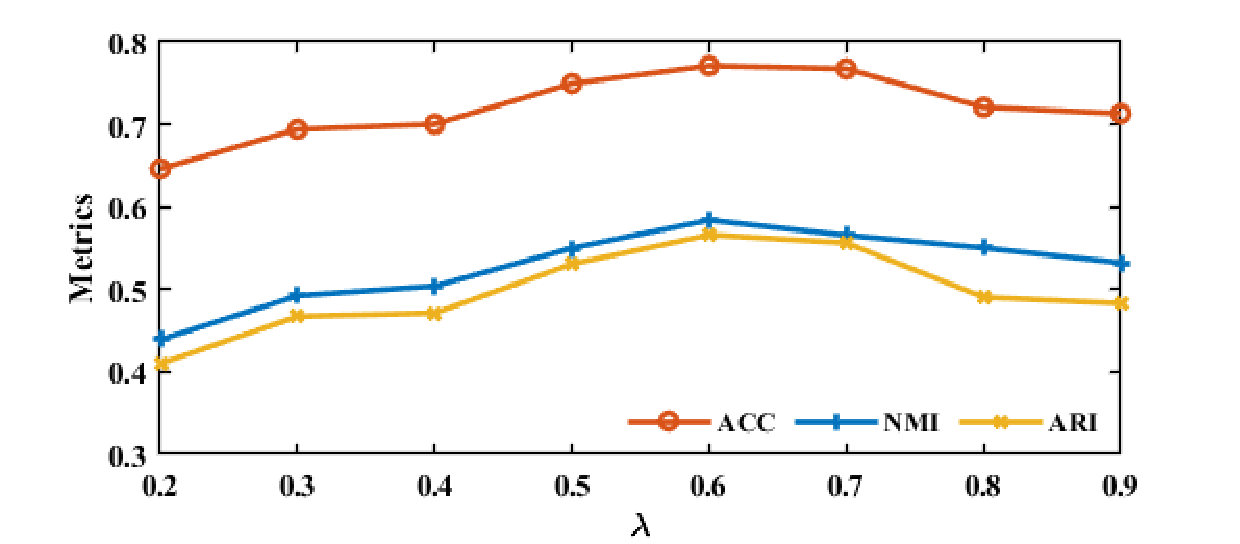}
	\caption{The parameter effect of the ratio of reserved nodes.}
\end{figure}

\subsection{Parameter sensitivity analysis}

\paragraph{Impact of hyperparameter $\lambda$} In our proposed model, there is a main hyperparameter $\lambda$ that balances the reconstruction loss and clustering loss. In the following, we will analyze the sensitivity of this parameter in the proposed method. In the experiments, we tune $\lambda$ from $\{10^{-2}, 10^{-1}, 1, 10, 10^2, 10^3, 10^4\}$ and the ACC and NMI of the final clustering results with different $\lambda$ are shown in Fig.~3. As shown in Fig.~3, when $\lambda$ is between $10$ and $100$, the corresponding evaluation metrics ACC and NMI will largely remain constant. 

\paragraph{Impact of ratio of reserved nodes} The current graph pooling methods usually set the ratio to 0.5. Since our goal is to learn the representations of nodes rather than representation of the graph, we do not use the common settings in our approach. We search it in the range of $\{0.2, 0.3, 0.4, 0.5, 0.6, 0.7, 0.8, 0.9\}$. And Fig.~4 shows the results of our method using different ratio of reserved nodes (taking CiteSeer as an example). It is observed that the promising performance could be expected when the ratio of reserved nodes is between 0.6 and 0.7. The reason we cannot choose a smaller ratio as other graph pooling methods is that dropping too many nodes can lead to larger errors when classifying all nodes. Hence the ratio of selected nodes of NCPool is set to 0.6 for all datasets.

\subsection{Model Analysis}

\begin{table}[t]\small
\caption{Clustering results of different graph embeddings}
\vspace{-1em}
\begin{center}
\resizebox{8.5cm}{!}{
\begin{tabular}{c|ccc|ccc|ccc}
\toprule
\multirow{2}{*}{Method} & \multicolumn{3}{c|}{Cora} & \multicolumn{3}{c|}{CiteSeer} & \multicolumn{3}{c}{PubMed} \\ 
                        & ACC & NMI & ARI & ACC & NMI & ARI & ACC & NMI & ARI  \\ 
\midrule
DGEN-G                  & 0.725&0.521&0.468&0.631&0.369&0.370&0.671&0.282&0.297\\
DGEN-L                  & 0.819&0.660&0.661&0.717&0.468&0.488&0.782&0.413&0.498\\
DGEN                    & 0.771&0.576&0.566&0.686&0.430&0.445&0.695&0.297&0.320 \\
\bottomrule
\end{tabular}}
\end{center}  
\end{table}

\begin{table}[t]\small
\caption{Results of different global nodes assignment methods}
\vspace{-1em}
\begin{center}
\resizebox{8.5cm}{!}{
\begin{tabular}{c|ccc|ccc|ccc}
\toprule
\multirow{2}{*}{Method} & \multicolumn{3}{c|}{Cora} & \multicolumn{3}{c|}{CiteSeer} & \multicolumn{3}{c}{PubMed} \\ 
                        & ACC & NMI & ARI & ACC & NMI & ARI & ACC & NMI & ARI  \\ 
\midrule
DGEN-a                  &0.737&0.556&0.482&0.667&0.391&0.408&0.674&0.285&0.293 \\
DGEN                    &0.771&0.576&0.566&0.686&0.430&0.445&0.695&0.297&0.320 \\
\bottomrule
\end{tabular}}
\end{center}  
\end{table}

\paragraph{Effectiveness of spurious edges removal} As mentioned before, spurious edges can weaken the discriminability of the learned node representations. To validate this, we compare the clustering results of the local node representations learned with or without spurious connections. Specifically, we compare the clustering results of the selected node representations(DGEN-s) and the representation of the corresponding nodes in the global graph embedding(DGEN-g). And the results are shown in Table~\uppercase\expandafter{\romannumeral5}. It demonstrates that removing spurious edges can effectively improve the clustering performance of the learned representations.

\paragraph{Validity of the step-wise clustering} In our model, a two step strategy is used to obtain the clustering assignment of all nodes. Besides the classification step, the clustering assignment of all nodes can also be obtained by the adjacency relations. In order to show the superiority of our final classification step, we compare these two methods and the results are reported in Table~\uppercase\expandafter{\romannumeral6}, where DGEN-a refers to the final clustering results obtained by the adjacency relations. Obviously, it is more efficient to use classification step to obtain the final clustering results because obtaining the final clustering results based on adjacency requires spurious edge information, which can undermine the clustering performance.

\section{Conclusions}

In this paper, we propose a dual graph embedding network(DGEN) to learn a more robust representation for graph clustering. Like most graph learning methods, we first learn the global graph embedding using the Global GAT Encoder. To reduce the effect of spurious edges on graph clustering, we design a novel neighbor cluster pooling(NCPool), which can select informative subset of nodes and the corresponding edges. The pooling is based on the assumption that if a node and its nearest neighboring node are close to the same clustering center, this node is an informative node and the corresponding edge can be considered as a cluster-friendly edge. The Local GAT Encoder is then utilized to learn local graph embedding on the selected nodes and edges. To train the whole model, we apply self-optimizing clustering loss and reconstruction loss to make the learned representation more suitable for clustering. Then, conventional clustering algorithm K-means is performed on the selected node representations so as to obtain the local clustering assignments. Consequently, we use the local clustering assignments as labels to train a classifier that enables us to obtain the final clustering assignments of all nodes. Evaluation on several benchmark datasets demonstrates the effectiveness of DGEN comparing with diverse baselines.

\ifCLASSOPTIONcaptionsoff
  \newpage
\fi



%
\bibliographystyle{IEEEtran}
\bibliography{Mybib}

\end{document}